\documentclass[12pt]{article}
\usepackage{amsmath,amssymb}
\makeatletter\@addtoreset {equation}{section}\makeatother

\newtheorem{theo}{Theorem}[section]
\newtheorem{lem}[theo]{Lemma}

\newtheorem{prop}[theo]{Proposition}

\newtheorem{rem}[theo]{Remark}
{\hspace*{\fill}$\rule{.3\baselineskip}{.35\baselineskip}$\end{trivlist}}

\newcommand{\R}{\mathbb{R}}

\renewcommand{\geq}{\geqslant}
\renewcommand{\leq}{\leqslant}

\renewcommand{\phi}{\varphi}
\newcommand{\be}{\begin{eqnarray}}
\newcommand{\ee}{\end{eqnarray}}

\newcommand{\eps}{\varepsilon}

\begin{document}

\title{\bf Expansion of the energy of the ground state of the
  Gross--Pitaevskii equation in the Thomas--Fermi limit}
\author{Cl\'ement Gallo\\
{\small Institut de Math\'ematiques et de Mod\'elisation de Montpellier, Universit\'e Montpellier II,}\\ 
{\small 34095 Montpellier, France}}
\date{\today}
\maketitle

\begin{abstract}
From the asymptotic expansion of the ground state of the
Gross--Pitaevskii equation in the Thomas--Fermi limit given by Gallo
and Pelinovsky \cite{GP}, we infer an
asymptotic expansion of the kinetic, potential and total energy of the
ground state. In particular, we give a rigorous proof of the expansion
of the kinetic energy calculated by Dalfovo, Pitaevskii and Stringari
\cite{DPS} in the case where the space dimension is 3. Moreover, we
calculate one more term in this expansion, and we generalize the
result to space dimensions 1 and 2.
\end{abstract}

\section{Introduction}
After recent experiments with Bose--Einstein condensates, new interest
has been stimulated in the Gross--Pitaevskii equation
with a harmonic potential, taken here in its adimensional form
\begin{equation}
\label{GP} i\eps u_t + \eps^2 \Delta u + (1 - |x|^2) u - |u|^2 u = 0,
\quad x \in \R^d, \quad t \in \R_+,
\end{equation}
where the space dimension $d$ is one, two or three, $u(t,x) \in \mathbb{C}$ describes the wave function of a repulsive
Bose gas, and $\eps$ is a small
parameter that corresponds to the Thomas--Fermi approximation of a
nearly compact atomic cloud \cite{Fermi,Thomas}.

\noindent A ground state of the Bose-Einstein condensate is a positive,
time-independent solution $u(t,x) = \eta_{\eps}(x)$ of the Gross--Pitaevskii equation (\ref{GP}).
Namely, $\eta_{\eps}: \mathbb{R}^d \mapsto \mathbb{R}$ satisfies the stationary
Gross--Pitaevskii equation
\begin{equation}
\label{stationaryGP}
\eps^2\Delta\eta_\eps(x)+(1-|x|^2)\eta_\eps(x)-\eta_\eps^3(x)=0,  \quad
x \in \R^d,
\end{equation}
$\eta_\eps(x) > 0$ for all $x \in \R^d$, and $\eta_\eps$ has a finite
energy $E_\eps(\eta_\eps)$, where $E_\eps$ is given by
$$E_\eps(u)=\int_{\R^d}\left(\eps^2|\nabla u|^2+(|x|^2-1)u^2+\frac{1}{2}u^4\right)dx.$$

\noindent In dimensions one, two and three, provided $\eps$ is sufficiently
small, existence and uniqueness of a radial ground state $\eta_{\eps}$
is known
\cite{IM,GP}. It is also well known \cite{IM,AAB} that $\eta_{\eps}(x)$ converges to $\eta_0(x)$ as $\eps \to 0$ for all
$x \in \R^2$, where $\eta_0$ is the so called Thomas--Fermi approximation 
\begin{equation}
\label{Thomas-Fermi}
\eta_0(x) = \left\{ \begin{array}{cl} (1 - |x|^2)^{1/2} & \text{for }  |x| < 1, \\
0 & \text{for }  |x| > 1. \end{array} \right.
\end{equation}

\noindent Several quantities such as the kinetic energy
of the ground state $$E_{k,\eps}(\eta_\eps)=\eps^2\int_{\R^d}|\nabla\eta_\eps|^2dx,$$
can not be accurately approximated just by replacing the ground state
$\eta_\eps$ by $\eta_0$, because of a logarithmic divergence at the
boundary region $|x|=1$. In \cite{DPS}, in the three-dimensional case,
Dalfovo, Pitaevskii and Stringari give the correct behavior of the
order parameter in the boundary region, which for instance provides
the first term in the asymptotic expansion of $E_{k,\eps}(\eta_\eps)$
close to $\eps=0$. The purpose of this work is to show how we can
obtain rigorously asymptotic expansions (theoretically at arbitrarily
high order of accuracy) of quantities such as the kinetic energy of
the ground state $E_{k,\eps}(\eta_\eps)$, its potential energy 
$$E_{p}(\eta_\eps)=\int_{\R^d}(|x|^2-1)\eta_\eps^2dx,$$
or its total energy $E_\eps(\eta_\eps)$. In particular, for $d=3$, we
give a rigorous proof of the expansion of the kinetic energy of the
ground state given in \cite{DPS}, we calculate one more term in this
expansion, and we generalize the result to one and two-dimensional
cases. The calculation of these expansions rely on the expansion of $\eta_{\eps}$ into powers of $\eps$ in the
limit $\eps\to 0$, which was established in
\cite{GP}. So, let us first summarize the main ideas which provide this
expansion of $\eta_{\eps}$ in \cite{GP}.
\noindent Since $\eta_\eps$ is radially symmetric, we can define
a function $\nu_\eps$ on $J_{\eps} := (-\infty,\eps^{-2/3}]$ by
\begin{equation}
\label{asymptotic-scaling}
\eta_\eps(x)=\eps^{1/3}\nu_\eps\left(\frac{1-|x|^2}{\eps^{2/3}}\right),
\quad x\in\R^d.
\end{equation}
Next, we rewrite equation (\ref{stationaryGP}) in terms of the new real variable $y = (1-|x|^2)/\eps^{2/3}$. It is equivalent for
$\eta_\eps$ to solve (\ref{stationaryGP}) and for $\nu_{\eps}$ to solve
the differential equation
\begin{equation}
\label{nu-eq}
4 (1 - \eps^{2/3} y) \nu_{\eps}''(y) - 2 \eps^{2/3}d \nu_{\eps}'(y) + y \nu_{\eps}(y) - \nu_{\eps}^3(y) = 0, \quad
y \in J_{\eps}.
\end{equation}
Let $N\geq 0$ be an integer. We look for $\nu_\eps$ using the form
\be\label{exp}
\nu_\eps(y)=\sum_{n=0}^N\eps^{2n/3}\nu_n(y)+\eps^{2(N+1)/3}R_{N,\eps}(y),
\quad y \in J_{\eps}.
\ee
Expansion (\ref{exp}) provides a solution of equation (\ref{nu-eq}) if $\{\nu_n\}_{0\leq
  n\leq N}$ and $R_{N,\eps}$ satisfy equations (\ref{P2-eq}), (\ref{nun-eq}) and
(\ref{r-eq}) below.
\begin{itemize}
\item $\nu_0$ solves the Painlev\'e-II equation
\begin{equation}
\label{P2-eq}
4 \nu_0''(y) + y \nu_0(y) - \nu^3_0(y) = 0, \quad y \in \mathbb{R},
\end{equation}
\item for $1\leq n\leq N$, $\nu_n$ solves
\be\label{nun-eq}
-4\nu_n''(y) + W_0(y)\nu_n(y) =F_n(y),\quad y\in \R,
\ee
where
$$
W_0(y) = 3 \nu_0^2(y) - y
$$
and
$$
F_n(y)=-\!\!\!\!\!\!\!\!\underset{\tiny{\begin{array}{c}n_1,n_2,n_3<n\\n_1+n_2+n_3=n\end{array}}}{\sum}\!\!\!\!\nu_{n_1}(y)\nu_{n_2}(y)\nu_{n_3}(y)-2d\nu_{n-1}'(y)-4y\nu_{n-1}''(y),$$
\item $R_{N,\eps}$ solves \be \label{r-eq}
-4(1-\eps^{2/3}y)R_{N,\eps}''+2\eps^{2/3}dR_{N,\eps}' + W_0 R_{N,\eps}=F_{N,\eps}(y,R_{N,\eps}),
\quad y \in J_{\eps},
\ee
where \be
\lefteqn{F_{N,\eps}(y,R)\ =\
-(4y\nu_N''+2d\nu_N')-\sum_{n=0}^{2N-1}\eps^{2n/3}\underset{\tiny{\begin{array}{c}n_1+n_2+n_3=n+N+1\\0\leq
      n_1,n_2,n_3\leq
      N\end{array}}}{\sum}\nu_{n_1}\nu_{n_2}\nu_{n_3}}\nonumber\\
&&-\left(3\sum_{n=1}^{2N}\eps^{2n/3}\underset{\tiny{\begin{array}{c}n_1+n_2=n\\0\leq
      n_1,n_2\leq N\end{array}}}{\sum}\nu_{n_1}\nu_{n_2}\right)R-
      \left(3\sum_{n=N+1}^{2N+1}\eps^{2n/3}\nu_{n-(N+1)}\right)R^2-\eps^{4(N+1)/3}R^3.\nonumber
\ee
\end{itemize}
Notice that for $0\leq n\leq N$, $\nu_n(y)$ is defined for all $y\in
\R$ and does not depend
on $\eps$, whereas $R_{N,\eps}(y)$ is only defined for $y\in J_{\eps}$.

In order to describe accurately the convergence of $\eta_\eps$ to
$\eta_0$ by the first term (corresponding to $n=0$) in expansion (\ref{exp}),
$\nu_0$ shall be chosen in such a way that
$$\eps^{1/3}\nu_0\left(\frac{1-x^2}{\eps^{2/3}}\right)\longrightarrow\left\{\begin{array}{ll}\sqrt{1-|x|^2}&{\rm
  if} |x|<1\\0&{\rm if} |x|>1,\end{array}\right.$$
which means that $\nu_0$ has to satisfy the asymptotic behaviour
\begin{eqnarray}\label{nu0asymp}
\nu_0(y) \sim y^{1/2} \quad \mbox{as} \quad y\to +\infty \quad
\mbox{and} \quad \nu_0(y)\to 0 \quad \mbox{as} \quad y\to-\infty.
\end{eqnarray}
The Painlev\'e-II equation is known to have a unique solution $\nu_0$
with the asymptotic behaviour (\ref{nu0asymp}). This is the so-called
Hastings-McLeod solution. Moreover, the behaviour of $\nu_0(y)$ as
$y\to \pm\infty$ has been studied in details, for instance in
\cite{HM}, \cite{M}, \cite{FIKN}. Some of its properties
are summarized in the next proposition.

\begin{prop}\label{proposition-Painleve}
$\ $ \cite{HM,M,FIKN}
The Painlev\'e-II equation (\ref{P2-eq}) admits a unique solution
$\nu_0\in \mathcal{C}^{\infty}(\R)$ which satisfies (\ref{nu0asymp}).
This solution $\nu_0$ is strictly increasing on $\R$. The behaviour of
$\nu_0$ as $y\to-\infty$ is described by
\be\label{asympnu0-}
\nu_0(y)=\frac{1}{\sqrt{\pi}(-y)^{-1/4}}\exp\left(-\frac{1}{3}(-y)^{3/2}\right)
\left(1+\mathcal{O}(|y|^{-3/4})\right)\underset{y\to-\infty}{\approx}0,
\ee whereas as $y\to +\infty$,
\be\label{asympnu0+}
\nu_0(y)\underset{y\to+\infty}{\approx}y^{1/2}\sum_{n=0}^{\infty}
\frac{b_n}{(2y)^{3n/2}}, \ee where $b_0=1$, $b_1=0$, and for $n
\geq 0$,
$$
b_{n+2}=4(9n^2-1)b_n-\frac{3}{2}\sum_{m=1}^{n+1}b_mb_{n+2-m}-\frac{1}{2}\sum_{l=1}^{n}\sum_{m=1}^{n+1-l}b_lb_m b_{n+2-l-m}.
$$
\end{prop}

Once $\nu_0$ has been chosen to be the Hastings-McLeod solution of the
Painlev\'e-II equation (\ref{P2-eq}), we show by induction on $n$ that
there is a unique way to construct the sequence $(\nu_n)_{n\geq
  1}\subset H^\infty(\R)$ such that (\ref{nun-eq}) is satisfied for
every $n\geq 1$. Moreover, like for $\nu_0$, the asymptotic behaviour
of the $\nu_n$'s can be precisely described, as it is shown in the
next proposition.

\begin{prop}\label{proposition-mun}
$\ $ \cite{GP}
For every $n\geq 1$, 
$$\nu_n(y) \underset{y\to +\infty}{\approx}
y^{\beta-2n}\sum_{m=0}^{\infty} g_{n,m} y^{-3m/2} \text{ for some } \{ g_{n,m}\}_{m \in \mathbb{N}},$$
and\hfill $\nu_n(y) \underset{y\to -\infty}{\approx} 0,\quad $\hfill\null\\
where \hfill$\beta=\left\{\begin{array}{ll} -5/2&\text{if } d=1,\\ 1/2 &\text{if } d=2,3.\end{array}\right.$\hfill\null
\end{prop}

\begin{rem}
The coefficients $g_{n,m}$ can be calculated explicitely by plugging
this expansion of $\nu_n$ into (\ref{nun-eq}). For instance,
$\nu_1(y)\underset{y\to +\infty}{\sim}\frac{5(7-d)}{4}y^{-9/2}$ if $d=1$,
whereas $\nu_1(y)\underset{y\to +\infty}{\sim}=\frac{1-d}{2}y^{-3/2}$
if $d=2,3$.
\end{rem}

Finally, we close the argument by constructing a remainder term
$R_{N,\eps}$ that solves equation (\ref{r-eq}). We also prove
suitable estimates on $R_{N,\eps}$ which ensure that the last term in
the right hand side of (\ref{exp}) is indeed small compared to the
other ones, and that the solution of the stationary Gross-Pitaevskii
equation (\ref{stationaryGP}) that we have constructed by
(\ref{asymptotic-scaling}), (\ref{exp}) and our choices of the
$\nu_n$'s and $R_{N,\eps}$ is indeed the unique ground state of
(\ref{stationaryGP}) (in particular, it
is positive). More precisely, we have the following result.

\begin{prop}\label{theorem-main}
$\ $ \cite{GP}
For every $N\geq 0$, there exists $\eps_N > 0$ and $C_N>0$ such that for every
$0<\eps<\eps_N$, there is a solution $R_{N,\eps}\in
\mathcal{C}^\infty\cap L^\infty(J_{\eps})$ of equation (\ref{r-eq}) with
$$
\|R_{N,\eps}\|_{L^\infty(J_{\eps})} \leq C_N,\quad \quad \int_{-\infty}^{\eps^{-2/3}}R_{N,\eps}(y)^2W_0(y)(1-\eps^{2/3}y)^{d/2-1}dy\leq C_N$$
$$\text{and}\quad
S_{N,\eps}:x\mapsto R_{N,\eps}
\left(\frac{1-|x|^2}{\eps^{2/3}}\right)\in H^2(\R^d),
$$
such that the unique radially symmetric ground state of equation (\ref{stationaryGP}) in
$L^2(\R^d)$ writes
\be\label{expetaeps}
\eta_\eps(x)=\eps^{1/3}\sum_{n=0}^N\eps^{2n/3}\nu_n\left(
\frac{1-|x|^2}{\eps^{2/3}}\right)+\eps^{2N/3+1}R_{N,\eps}
\left(\frac{1-|x|^2}{\eps^{2/3}}\right),\quad x\in \R^d.
\ee
\end{prop}

\begin{rem}
The estimate on the $L^\infty$ norm of $R_{N,\eps}$ written in
\cite{GP} was $\|R_{N,\eps}\|_{L^\infty(J_{\eps})} \leq
C_N\eps^{-(d-1)/3}$. The estimate we have in Proposition
\ref{theorem-main} above is a direct consequence of this inequality
with $R_{N,\eps}$ replaced by $R_{N+1,\eps}$, taking into account that
$d\leq 3$, $R_{N,\eps}=\nu_{N+1}+\eps^{2/3}R_{N+1,\eps}$ and $\nu_{N+1}\in L^\infty(\R)$. The other estimate on $R_{N,\eps}$ is a byproduct of the proof of Proposition \ref{theorem-main} given in \cite{GP}. Indeed, $R_{N,\eps}$ is obtained there thanks to a fix point argument in a ball $B_{H_\eps^1}(R_{N,\eps}^0,C\eps^{2/3})$, where $\|R_{N,\eps}^0\|_{H_\eps^1}\lesssim 1$ and 
$$\|u\|_{H_\eps^1}^2=\int_{-\infty}^{\eps^{-2/3}}\left(4(1-\eps^{2/3}y)^{d/2}|u'|^2+(1-\eps^{2/3}y)^{d/2-1}W_0(y)u(y)^2\right)dy.$$
\end{rem}

The asymptotic expansion of $\eta_\eps$ given by
(\ref{asymptotic-scaling})-(\ref{exp}) as well as the precise
description of the behaviour of $\nu_n(y)$ as $y\to\pm\infty$ (for
$n\geq 0$) given in Propositions \ref{proposition-Painleve} and
\ref{proposition-mun}, enable us to calculate expansions for kinetic,
potential and full energy of the ground state. Concerning the full
energy, we get the following expansion.

\begin{theo}\label{expansion}
For $d=1$,
\begin{eqnarray*}
E_\eps(\eta_\eps)&=&-\frac{8}{15}-\frac{2}{3}\eps^2\ln\eps+\left[\int_0^1\frac{(1-t)^{-\frac{1}{2}}-1}{t}dt-\frac{1}{2}\int_0^{+\infty}\left(\nu_0(y)^4-y^2+\frac{2}{y}\mathbf{1}_{\{y\geq
    1\}}\right)dy\right]\eps^2\\
&&\hspace{-1.5cm}+\left[-\frac{1}{2}-\frac{1}{4}\int_{-\infty}^{+\infty}y\left(\nu_0(y)^4-y_+^2+\frac{2}{y}\mathbf{1}_{\{y\geq
    1\}}\right)dy-2\int_{-\infty}^{+\infty}\nu_0(y)^3\nu_1(y)dy\right]\eps^{8/3}+O(\eps^3).
\end{eqnarray*}
For $d=2$,
\begin{eqnarray*}
E_\eps(\eta_\eps)&=&-\frac{\pi}{6}-\frac{2\pi}{3}\eps^2\ln\eps+\left[-\frac{\pi}{2}\int_0^{+\infty}\left(\nu_0(y)^4-y^2+\frac{2}{y}\mathbf{1}_{\{y\geq
    1\}}\right)dy+\pi\right]\eps^2\\
&&-2\pi\left[\int_{-\infty}^{+\infty}\left(\nu_0(y)^3\nu_1(y)+\frac{1}{2}\mathbf{1}_{\{y\geq
    0\}}\right)dy\right]\eps^{8/3}+O(\eps^3).
\end{eqnarray*}
For $d=3$,
\begin{eqnarray*}
\lefteqn{E_\eps(\eta_\eps)=-\frac{16\pi}{105}-\frac{4\pi}{3}\eps^2\ln\eps}\\
&&+\left[2\pi\int_0^1\frac{(1-t)^{\frac{1}{2}}-1}{t}dt-\pi\int_0^{+\infty}\left(\nu_0(y)^4-y^2+\frac{2}{y}\mathbf{1}_{\{y\geq
    1\}}\right)dy+\frac{8}{3}\right]\eps^2\\
&&\hspace{-0.8cm}+\left[\pi+\frac{\pi}{2}\int_{-\infty}^{+\infty}y\left(\nu_0(y)^4-y_+^2+\frac{2}{y}\mathbf{1}_{\{y\geq
    1\}}\right)dy-4\pi\int_{-\infty}^{+\infty}\left(\nu_0(y)^3\nu_1(y)+\mathbf{1}_{\{y\geq 0
    \}}\right)dy\right]\eps^{8/3}+O(\eps^3).
\end{eqnarray*}
\end{theo}

\begin{rem}
Note that the notation $\nu_1$ does not represent the same function in
the three expansions given in Theorem \ref{expansion}. Indeed, the
equation (\ref{nun-eq}) satisfied by $\nu_1$ depends on the dimension
$d$ through $F_1$.
\end{rem}

The rest of the paper is organized as follows. In section \ref{sec2}
we calculate asymptotic expansions of $E_\eps(\eta_\eps)$ and prove
Theorem \ref{expansion}. In section \ref{sec3} we calculate the
asymptotic expansion of the potential energy $E_p(\eta_\eps)$. In
section \ref{sec4}, we deduce the expansion of the kinetic energy 
from the results of the two previous section, and we rediscover the
expansion found by Dalfovo, Pitaevskii and Stringari in \cite{DPS} on
a formal level. In the appendix, we prove a key lemma which is used on
many occasions in the calculation.

\section{Expansion of $E_\eps(\eta_\eps)$}\label{sec2}
We are interested here in the behaviour of
$E_\eps(\eta_\eps)$ as $\eps \to 0$. First, if we multiply
(\ref{stationaryGP}) by $\eta_\eps$ and sum over $\R^d$, we get
\begin{eqnarray}\label{depart}
E_\eps(\eta_\eps)=-\frac{1}{2}\int_{\R^d}\eta_\eps(x)^4dx.
\end{eqnarray}
From the convergence of $\eta_\eps$ to $\eta_0$ in $L^p(\R^d)$ (for
any $p\in [1,+\infty]$) as $\eps\to 0$ \cite{GP}, we already know
$$E_\eps(\eta_\eps)\underset{\eps\to
  0}{\longrightarrow}-\frac{1}{2}\int_{\R^d}\eta_0(x)^4dx=\left\{\begin{array}{ll}-8/15&\text{if
} d=1\\-\pi/6&\text{if
} d=2\\-16\pi/105&\text{if
} d=3.\end{array}\right.$$

Next, we calculate some correction terms in the asymptotic
expansion of $E_\eps(\eta_\eps)$ as $\eps\to 0$. From (\ref{depart}),
(\ref{asymptotic-scaling}) and (\ref{exp}) we infer

\begin{eqnarray*}
\lefteqn{E_\eps(\eta_\eps)+\frac{1}{2}\int_{\R^d}\eta_0(x)^4dx}\\
&=&-\frac{1}{2}\int_{\R^d}\left(\eta_\eps(x)^4-\eta_0(x)^4\right)dx\\
&=&-\frac{1}{2}\int_{\R^d}\eps^{4/3}\left(\nu_\eps\left(\frac{1-|x|^2}{\eps^{2/3}}\right)^4-\sqrt{\left(\frac{1-|x|^2}{\eps^{2/3}}\right)_+}^4\right)dx\\
&=&-\frac{\eps^{4/3}}{2}\left|\mathbb{S}^{d-1}\right|\int_0^{+\infty}\left(\nu_\eps\left(\frac{1-r^2}{\eps^{2/3}}\right)^4-\sqrt{\left(\frac{1-r^2}{\eps^{2/3}}\right)_+}^4\right)r^{d-1}dr\\
&=&-\frac{\eps^{2}}{4}\left|\mathbb{S}^{d-1}\right|\int_{-\infty}^{\eps^{-2/3}}\left(\nu_\eps(y)^4-\sqrt{y_+}^4\right)(1-\eps^{2/3}y)^{d/2-1}dy\\
&=&-\frac{\eps^{2}}{4}\left|\mathbb{S}^{d-1}\right|\int_{-\infty}^{\eps^{-2/3}}\left((\nu_0(y)+\eps^{2/3}R_{0,\eps}(y))^4-\sqrt{y_+}^4\right)(1-\eps^{2/3}y)^{d/2-1}dy\\
&=&-\frac{\eps^{2}}{4}\left|\mathbb{S}^{d-1}\right|\sum_{j=1}^6I_j,
\end{eqnarray*}
where
\begin{eqnarray*}
I_1 & =& \int_{-\infty}^{\eps^{-2/3}}-\frac{2}{y}\mathbf{1}_{\{y\geq
    1\}}(1-\eps^{2/3}y)^{d/2-1}dy,\\
I_2&=&\int_{-\infty}^{\eps^{-2/3}}\left(\nu_0(y)^4-\sqrt{y_+}^4+\frac{2}{y}\mathbf{1}_{\{y\geq
    1\}}\right)(1-\eps^{2/3}y)^{d/2-1}dy,\\
I_3&=&4\eps^{2/3}\int_{-\infty}^{\eps^{-2/3}}\nu_0(y)^3R_{0,\eps}(y)(1-\eps^{2/3}y)^{d/2-1}dy,\\
I_4&=&6\eps^{4/3}\int_{-\infty}^{\eps^{-2/3}}\nu_0(y)^2R_{0,\eps}(y)^2(1-\eps^{2/3}y)^{d/2-1}dy,\\
I_5&=&4\eps^{2}\int_{-\infty}^{\eps^{-2/3}}\nu_0(y)R_{0,\eps}(y)^3(1-\eps^{2/3}y)^{d/2-1}dy,\\
I_6&=&\eps^{8/3}\int_{-\infty}^{\eps^{-2/3}}R_{0,\eps}(y)^4(1-\eps^{2/3}y)^{d/2-1}dy.
\end{eqnarray*}

Next, we give an asymptotic expansion as $\eps\to 0$ of each of the
$I_j$'s, with a $O(\eps)$ remainder. For this purpose, the following
lemma, which is proved in the appendix, will be convenient.

\begin{lem}\label{lemg}
Let $g:\R\mapsto\R$ be a bounded function, such that
$g(y)\underset{y\to-\infty}{=}O(\exp(y))$, and
$g(y)\underset{y\to+\infty}{=}O(y^{-\alpha})$, where
$\alpha\in\R$. Then
\be
\lefteqn{\int_{-\infty}^{\eps^{-2/3}}g(y)(1-\eps^{2/3}y)^{d/2-1}dy}\\
&&=\left\{\begin{array}{ll}
O(\eps^{-1/3})&
\text{if } \alpha\geq 1/2\\
\int_{-\infty}^{+\infty}g(y)dy+O(\eps^{1/3})&
\text{if } \alpha\geq 3/2\\
\int_{-\infty}^{+\infty}g(y)dy-\left(\frac{d}{2}-1\right)\eps^{2/3}\int_{-\infty}^{+\infty}yg(y)dy+O(\eps)&
\text{if } \alpha\geq 5/2.
\end{array}\right.
\ee
\end{lem}

\paragraph{Expansion of $I_1$.} From the change of variable
$t=\eps^{2/3}y$, we get
\begin{eqnarray*}
I_1 &=& -2\int_{1}^{\eps^{-2/3}}(1-\eps^{2/3}y)^{d/2-1}\frac{dy}{y}\\
 &=& -2\int_{\eps^{2/3}}^1(1-t)^{d/2-1}\frac{dt}{t}\\
 &=&
-2\int_{\eps^{2/3}}^1\frac{dt}{t}-2\int_0^1\frac{(1-t)^{d/2-1}-1}{t}dt+2\int_0^{\eps^{2/3}}\frac{(1-t)^{d/2-1}-1}{t}dt\\
 &=&\frac{4}{3}\ln\eps-2\int_0^1\frac{(1-t)^{d/2-1}-1}{t}dt-2\left(\frac{d}{2}-1\right)\eps^{2/3}+O(\eps^{4/3}).
\end{eqnarray*}

\paragraph{Expansion of $I_2$.} We apply Lemma \ref{lemg} to the function
$$g_0(y):=\nu_0(y)^4-\sqrt{y_+}^4+\frac{2}{y}\mathbf{1}_{\{y\geq
    1\}}.$$
Note that Proposition \ref{proposition-Painleve} yields $g_0(y)\underset{y\to-\infty}{=}O(\exp(y))$, and
$g_0(y)\underset{y\to+\infty}{=}O(y^{-5/2})$. Thus,
\begin{eqnarray*}
I_2&=&\int_{-\infty}^{\eps^{-2/3}}g_0(y)(1-\eps^{2/3}y)^{d/2-1}dy=\int_{-\infty}^{+\infty}g_0(y)dy-\left(\frac{d}{2}-1\right)\eps^{2/3}\int_{-\infty}^{+\infty}yg_0(y)dy+O(\eps).
\end{eqnarray*}

\paragraph{Expansion of $I_3$.} From (\ref{exp}), we get
\begin{eqnarray*}
I_3&=&4\eps^{2/3}\int_{-\infty}^{\eps^{-2/3}}\nu_0(y)^3R_{0,\eps}(y)(1-\eps^{2/3}y)^{d/2-1}dy\\
&=&4\eps^{2/3}\int_{-\infty}^{\eps^{-2/3}}\nu_0(y)^3\left(\sum_{j=0}^k\eps^{2j/3}\nu_{j+1}(y)+\eps^{2(k+1)/3}R_{k+1,\eps}(y)\right)(1-\eps^{2/3}y)^{d/2-1}dy.\\
\end{eqnarray*}
From Propositions \ref{proposition-Painleve} and
\ref{proposition-mun}, we have
$$\nu_0(y)\underset{y\to +\infty}{=}y^{1/2}-\frac{1}{2}y^{-5/2}+O(y^{-4}),$$
$$\nu_1(y)\underset{y\to +\infty}{=}\left\{\begin{array}{ll}\frac{1-d}{2}y^{-3/2}+O(y^{-3}) &
\text{if } d=2,3\\ O(y^{-9/2})&\text{if } d=1,\end{array}\right.$$ 
and  $\nu_0(y), \nu_1(y)\underset{y\to +\infty}{\approx}0$. In particular, for $d=2,3$,
$$\nu_0(y)^3\nu_1(y)\underset{y\to +\infty}{=}\left(y^{1/2}+O(y^{-5/2})\right)^3\left(\frac{1-d}{2}y^{-3/2}+O(y^{-3})\right)=\frac{1-d}{2}+O(y^{-3/2}),$$
a result which also holds for $d=1$. Thus, Lemma \ref{lemg} implies
\begin{eqnarray}\label{i31}
\lefteqn{4\eps^{2/3}\int_{-\infty}^{\eps^{-2/3}}\nu_0(y)^3\nu_1(y)(1-\eps^{2/3}y)^{d/2-1}dy}\nonumber\\
&=&4\eps^{2/3}\int_{-\infty}^{\eps^{-2/3}}g_1(y)(1-\eps^{2/3}y)^{d/2-1}dy-4\eps^{2/3}\int_0^{\eps^{-2/3}}\frac{d-1}{2}(1-\eps^{2/3}y)^{d/2-1}dy\nonumber\\
&=&4\eps^{2/3}\int_{-\infty}^{\eps^{-2/3}}g_1(y)(1-\eps^{2/3}y)^{d/2-1}dy-4\frac{d-1}{d}\nonumber\\
&=&4\frac{1-d}{d}+4\eps^{2/3}\int_{-\infty}^{+\infty}g_1(y)dy+O(\eps),
\end{eqnarray}
where
$$g_1(y)=\nu_0(y)^3\nu_1(y)+\frac{d-1}{2}\mathbf{1}_{\{y\geq
  0\}}\underset{y\to +\infty}{=}O(y^{-3/2}).$$
Next Proposition \ref{proposition-mun} provides, for $j\geq 1$,
$\nu_{j+1}(y)\underset{y\to+\infty}{=}O(y^{-7/2})$, and therefore
$\nu_0(y)^3\nu_{j+1}(y)=O(y^{-2})$. Thus, Lemma \ref{lemg} implies
\begin{eqnarray}\label{i32}
4\eps^{2(1+j)/3}\int_{-\infty}^{\eps^{-2/3}}\nu_0(y)^3\nu_{j+1}(y)(1-\eps^{2/3}y)^{d/2-1}dy&=&O(\eps^{2(1+j)/3})=O(\eps^{4/3}).
\end{eqnarray}
Finally, if $\eps\leq 1$,
\begin{eqnarray}\label{i33}
\lefteqn{\left|4\eps^{2/3}\int_{-\infty}^{\eps^{-2/3}}\nu_0(y)^3\eps^{2(k+1)/3}R_{k+1,\eps}(y)(1-\eps^{2/3}y)^{d/2-1}dy\right|}\nonumber\\
&\leq&4\eps^{2(k+2)/3}\|R_{k+1,\eps}\|_{L^\infty(J_\eps)}\int_{-\infty}^{\eps^{-2/3}}\nu_0(y)^3(1-\eps^{2/3}y)^{d/2-1}dy\nonumber\\
&\lesssim
&\eps^{2(k+2)/3}\left(\int_{-\infty}^{0}\nu_0(y)^3(1+|y|)^{1/2}dy+(\eps^{-2/3})^{3/2}\int_0^{\eps^{-2/3}}(1-\eps^{2/3}y)^{d/2-1}dy\right)\nonumber\\
&\lesssim&\eps^{2(k+2)/3}(1+\eps^{-5/3})=O(\eps),
\end{eqnarray}
provided $k\geq 2$. With such a choice of $k$, the combination of estimates (\ref{i31}), (\ref{i32}) and (\ref{i33}) yields 
\begin{eqnarray*}
I_3&=&4\frac{1-d}{d}+4\eps^{2/3}\int_{-\infty}^{+\infty}g_1(y)dy+O(\eps).
\end{eqnarray*}

\paragraph{Estimates on $I_4$,$I_5$,$I_6$.}
As mentionned in Proposition \ref{proposition-Painleve}, $\nu_0$ is an increasing function on $\R$, $\nu_0(y)\underset{y\to -\infty}{\longrightarrow}0$ and $\nu_0(y)\underset{y\to +\infty}{\sim}y^{1/2}$. On the other side, it is proved in \cite{GP} that $W_0(y)=3\nu_0(y)^2-y>0$ for $y\in \R$. Thus, there exists $C>0$ such that for every $y\in \R$,
$$\max(\nu_0(y)^2,\nu_0(y),1)\leq CW_0(y).$$
Thus, it follows from the estimates on $R_{0,\eps}$ stated in Theorem \ref{theorem-main} that
$$I_4=O(\eps^{4/3}),\quad I_5=O(\eps^{2}),\quad I_6=O(\eps^{8/3}). $$

Combining the asymptotic expansions of all of the $I_j$'s, we obtain finally

\begin{eqnarray*}
\lefteqn{E_\eps(\eta_\eps)+\frac{1}{2}\int_{\R^d}\eta_0(x)^4dx}\\
&=&-\frac{\eps^{2}}{4}\left|\mathbb{S}^{d-1}\right|\left[\frac{4}{3}\ln\eps+\left(4\frac{1-d}{d}-2\int_0^1\frac{(1-t)^{d/2-1}-1}{t}dt+\int_{-\infty}^{+\infty}g_0(y)dy\right)\right.\\
&&\left.
+\left((1-\frac{d}{2})(2+\int_{-\infty}^{+\infty}yg_0(y)dy)+4\int_{-\infty}^{+\infty}g_1(y)dy\right)\eps^{2/3}+O(\eps)\right].
\end{eqnarray*}

\section{Expansion of $E_p(\eta_\eps)$}\label{sec3}
A calculation similar to the one we made to compute
$E_\eps(\eta_\eps)+\frac{1}{2}\|\eta_0\|_{L^4(\R^d)}^4$ gives
\begin{eqnarray*}
\lefteqn{E_p(\eta_\eps)-\int_{\R^d}\left(|x|^2-1\right)\eta_0(x)^2dx}\\
&=&-\frac{\eps^{2}}{2}\left|\mathbb{S}^{d-1}\right|\int_{-\infty}^{\eps^{-2/3}}y\left((\nu_0(y)+\eps^{2/3}R_{0,\eps}(y))^2-\sqrt{y_+}^2\right)(1-\eps^{2/3}y)^{d/2-1}dy\\
&=&-\frac{\eps^{2}}{2}\left|\mathbb{S}^{d-1}\right|\sum_{j=1}^4J_j,
\end{eqnarray*}
where
\begin{eqnarray*}
J_1 & =& \int_{-\infty}^{\eps^{-2/3}}-\frac{1}{y}\mathbf{1}_{\{y\geq
    1\}}(1-\eps^{2/3}y)^{d/2-1}dy,\\
J_2&=&\int_{-\infty}^{\eps^{-2/3}}\left(y(\nu_0(y)^2-y_+)+\frac{1}{y}\mathbf{1}_{\{y\geq
    1\}}\right)(1-\eps^{2/3}y)^{d/2-1}dy,\\
J_3&=&2\eps^{2/3}\int_{-\infty}^{\eps^{-2/3}}y\nu_0(y)R_{0,\eps}(y)(1-\eps^{2/3}y)^{d/2-1}dy,\\
J_4&=&\eps^{4/3}\int_{-\infty}^{\eps^{-2/3}}yR_{0,\eps}(y)^2(1-\eps^{2/3}y)^{d/2-1}dy.
\end{eqnarray*}
The method used to calculate asymptotic expansions of the $J_j$'s for
$j=1,2,3,4$ is
similar to the one we used for the $I_j$'s. More precisely,

\paragraph{Expansion of $J_1$.} 
\begin{eqnarray*}
J_1 =\frac{1}{2} I_1=\frac{2}{3}\ln\eps-\int_0^1\frac{(1-t)^{d/2-1}-1}{t}dt-\left(\frac{d}{2}-1\right)\eps^{2/3}+O(\eps^{4/3}).
\end{eqnarray*}

\paragraph{Expansion of $J_2$.} 
From Lemma \ref{lemg},
\begin{eqnarray*}
J_2&=&=\int_{-\infty}^{+\infty}g_2(y)dy-\left(\frac{d}{2}-1\right)\eps^{2/3}\int_{-\infty}^{+\infty}yg_2(y)dy+O(\eps),
\end{eqnarray*}
where
$$g_2(y)=y(\nu_0(y)^2-y_+)+\frac{1}{y}\mathbf{1}_{\{y\geq
  1\}}\underset{y\to +\infty}{=} O(y^{-5/2}).$$

\paragraph{Expansion of $J_3$.} We infer from Propositions
(\ref{proposition-Painleve}) and (\ref{proposition-mun}) that
$$y\nu_0(y)\nu_1(y)\underset{y\to
  +\infty}{=}\frac{1-d}{2}+O(y^{-3/2}).$$
We put
$$g_3(y):= y\nu_0(y)\nu_1(y)+\frac{d-1}{2}\mathbf{1}_{\{y\geq
  0\}}\underset{y\to
  +\infty}{=}O(y^{-3/2}).$$
Like in the calculation of $I_3$, we notice that for every $j\geq 1$, $y\nu_0(y)\nu_{j+1}(y)\underset{y\to
  +\infty}{=}O(y^{-2})$, and we deduce
$$J_3=2\frac{1-d}{d}+2\eps^{2/3}\int_{-\infty}^{+\infty}g_3(y)dy+O(\eps).$$

\paragraph{Expansion of $J_4$.} Like in the estimate on $I_4$, we
have 
$$J_4=O(\eps^{4/3}).$$

As a conclusion, we get

\begin{eqnarray*}
\lefteqn{E_p(\eta_\eps)+\int_{\R^d}\eta_0(x)^4dx}\\
&=&-\frac{\eps^{2}}{2}\left|\mathbb{S}^{d-1}\right|\left[\frac{2}{3}\ln\eps+\left(2\frac{1-d}{d}-\int_0^1\frac{(1-t)^{d/2-1}-1}{t}dt+\int_{-\infty}^{+\infty}g_2(y)dy\right)\right.\\
&&\left.
+\left((1-\frac{d}{2})(1+\int_{-\infty}^{+\infty}yg_2(y)dy)+2\int_{-\infty}^{+\infty}g_3(y)dy\right)\eps^{2/3}+O(\eps)\right].
\end{eqnarray*}

\section{Expansion of $E_{k,\eps}(\eta_\eps)$}\label{sec4}
Let us multiply (\ref{stationaryGP}) by $\eta_\eps$ and sum over $\R^d$. We get
\begin{eqnarray*}
E_{k,\eps}(\eta_\eps)&=&2E_{\eps}(\eta_\eps)-E_p(\eta_\eps)
\end{eqnarray*}
Thus, the asymptotic expansions of $E_{\eps}(\eta_\eps)$ and $E_p(\eta_\eps)$ obtained in the previous sections ensure
\begin{eqnarray*}
\lefteqn{E_{k,\eps}(\eta_\eps)}\\&=&-\frac{\eps^{2}}{2}\left|\mathbb{S}^{d-1}\right|\left[\frac{2}{3}\ln\eps+\left(2\frac{1-d}{d}-\int_0^1\frac{(1-t)^{d/2-1}-1}{t}dt+\int_{-\infty}^{+\infty}(g_0(y)-g_2(y))dy\right)\right.\\
&&\left.
+\left((1-\frac{d}{2})(1+\int_{-\infty}^{+\infty}y(g_0(y)-g_2(y))dy)+\int_{-\infty}^{+\infty}(4g_1(y)-2g_3(y))dy\right)\eps^{2/3}+O(\eps)\right]\\
&=&-\frac{\eps^{2}}{2}\left|\mathbb{S}^{d-1}\right|\left[\frac{2}{3}\ln\eps+\left(2\frac{1-d}{d}-\int_0^1\frac{(1-t)^{d/2-1}-1}{t}dt\right.\right.\\
&&\left.\left.+\int_{-\infty}^{+\infty}(\nu_0(y)^2(\nu_0(y)^2-y)+\frac{1}{y}\mathbf{1}_{\{y\geq
    1\}})dy\right)\right.\\
&&\left.+\left((1-\frac{d}{2})(\int_{-\infty}^{+\infty}\left(y\nu_0(y)^2(\nu_0(y)^2-y)+\mathbf{1}_{\{y\geq
    0\}}\right)dy)\right.\right.\\
&&\left.\left.+2\int_{-\infty}^{+\infty}\left((2\nu_0(y)^2-y)\nu_0(y)\nu_1(y)+(d-1)\mathbf{1}_{\{y\geq 0\}}\right)dy\right)\eps^{2/3}+O(\eps)\right].
\end{eqnarray*}

In \cite{DPS}, the author study the kinetic energy
$$E_{kin}=\frac{\hbar^2}{2m}\int_{\R^3}\left|\nabla\psi(\mathbf{r})\right|^2d\mathbf{r}$$
of the ground state $\psi$ of the Gross--Pitaevskii equation with an isotropic harmonic trap
\begin{eqnarray}\label{GPphys}
\frac{\hbar^2}{2m}\Delta\psi(\mathbf{r})+\left(\mu-\frac{1}{2}m\omega_{HO}^2|\mathbf{r}|^2\right)\psi(\mathbf{r})-\frac{4\pi\hbar^2 a}{m}|\psi(\mathbf{r})|^2\psi(\mathbf{r})=0,
\end{eqnarray}
where $\mu>0$ is a chemical potential, $a>0$ is the scattering
length. Like in \cite{DPS}, we denote  by $R$ the radius of the condensate, defined by
$$\mu=\frac{1}{2}m\omega_{HO}^2R^2,$$
and we introduce the harmonic oscillator length
$$a_{HO}=\left(\frac{\hbar}{m\omega_{HO}}\right)^{1/2},$$
as well as the maximal value $\alpha$ of the wave function $\psi_{TF}$ of the Thomas-Fermi approximation:
$$\alpha=\frac{R}{(8\pi a_{HO}^4a)^{1/2}},\quad
\psi_{TF}(\mathbf{r})=\alpha\left(1-\frac{\mathbf{r}^2}{R^2}\right)\text{
  if } |\mathbf{r}|\leq 1.$$
Then, the total number of particles is
$$N=\int_{|\mathbf{r}|\leq 1}\psi_{TF}(\mathbf{r})^2d\mathbf{r}=\frac{R^5}{15aa_{HO}^4}.$$
The change of variables
$$\psi(\mathbf{r})=\alpha u(\mathbf{r}/R),$$
maps the ground state $\psi$ of (\ref{GPphys}) to the solution $u=\eta_\eps$
of (\ref{stationaryGP}), where
$$\eps=\frac{a_{HO}^2}{R^2}.$$
Let us now use our expansion of $E_{k,\eps}(\eta_\eps)$ to calculate
the expansion of $E_{kin}$.
\begin{eqnarray*}
E_{kin}
&=&\frac{\hbar^2}{2m}\alpha^2R\int_{\R^3}\left|\nabla\eta_\eps(x)\right|^2dx\\
&=&\frac{\hbar^2}{2m}\frac{R^3}{8\pi
  a_{HO}^4a}\eps^{-2}E_{k,\eps}(\eta_\eps)\\
&=&-\frac{\hbar^2R^3}{8m
  a_{HO}^4a}\left[\frac{2}{3}\ln\frac{a_{HO}^2}{R^2}-\frac{4}{3}-\int_0^1\frac{(1-t)^{1/2}-1}{t}dt\right.\\
&&\left.+\int_{-\infty}^{+\infty}(g_0(y)-g_2(y))dy+O\left(\left(\frac{a_{HO}}{R}\right)^{4/3}\right)\right]\\
&=&-\frac{5N\hbar^2}{2mR^2}\left[\ln\frac{a_{HO}}{R}+\frac{1}{2}-\frac{3}{2}\ln
  2+\frac{3}{4}\int_{-\infty}^{+\infty}(\nu_0(y)^4-y\nu_0(y)^2\right.\\
&&\left.+\frac{\mathbf{1}_{\{y\geq
    1\}}}{y})dy+O\left(\left(\frac{a_{HO}}{R}\right)^{4/3}\right)\right].\\
\end{eqnarray*}
Moreover, taking into account the properties of the solution $\nu_0$
of the Painlev\'e mentioned in Proposition
\ref{proposition-Painleve}, we infer
$$\frac{1}{4}\int_{-\infty}^{+\infty}\left(\nu_0(y)^4-y\nu_0(y)^2+\frac{\mathbf{1}_{\{y\geq
    1\}}}{y}\right)dy=\frac{1}{2}-\int_{-\infty}^{+\infty}\left(\nu_0'(y)^2-\frac{\mathbf{1}_{\{y\geq
    1\}}}{4y}\right)dy.$$
In order to compare our result with the one in \cite{DPS}, we
introduce the function $\phi(\xi)=2^{-1/3}\nu_0(-2^{-2/3}\xi)$, which
is the solution to 
$$\phi''-\xi\phi-\phi^3=0,\quad \phi(\xi)\underset{\xi\to-\infty}{\sim}\sqrt{-\xi},\quad\phi(\xi)\underset{\xi\to+\infty}{\longrightarrow}0.$$
Thus,
\begin{eqnarray*}
E_{kin}
&=&-\frac{5N\hbar^2}{2mR^2}\left[\ln\frac{a_{HO}}{R}+2-\frac{3}{2}\ln
  2-3\int_{-\infty}^{+\infty}\left(\nu_0'(y)^2-\frac{\mathbf{1}_{\{y\geq
    1\}}}{4y}\right)dy+O\left(\left(\frac{a_{HO}}{R}\right)^{4/3}\right)\right]\\
&=&-\frac{5N\hbar^2}{2mR^2}\left[\ln\frac{a_{HO}}{R}+2-\frac{3}{2}\ln
  2-3\int_{-\infty}^{+\infty}\left(\phi'(\xi)^2+\frac{\mathbf{1}_{\{\xi\leq
    -2^{-2/3}\}}}{4\xi}\right)d\xi+O\left(\left(\frac{a_{HO}}{R}\right)^{4/3}\right)\right]\\
&=&\frac{5N\hbar^2}{2mR^2}\left[\ln\frac{R}{a_{HO}}-2+\frac{3}{2}\ln
  2\right.\\
&&\left.+3\underset{A\to+\infty}{\lim}\left(\int_{-A}^{+\infty}\phi'(\xi)^2d\xi+\frac{1}{4}\ln2^{-2/3}-\frac{1}{4}\ln
  A\right)+O\left(\left(\frac{a_{HO}}{R}\right)^{4/3}\right)\right]\\
&=&\frac{5N\hbar^2}{2mR^2}\left[\ln\frac{R}{a_{HO}}-2+\ln
  2+3\left(C+\frac{1}{4}\ln2\right)+O\left(\left(\frac{a_{HO}}{R}\right)^{4/3}\right)\right]\\
&=&\frac{5N\hbar^2}{2mR^2}\left[\ln\frac{R}{a_{HO}}-2+\frac{7}{4}\ln
  2+3C+O\left(\left(\frac{a_{HO}}{R}\right)^{4/3}\right)\right],\\
\end{eqnarray*}
where like in \cite{DPS}, we have denoted 
$$C=\underset{A\to+\infty}{\lim}\left(\int_{-A}^{+\infty}\phi'(\xi)^2d\xi-\frac{1}{4}\ln A\right)-\frac{1}{4}\ln2.$$

\section{Appendix: Proof of Lemma \ref{lemg}}
If $\alpha\geq 1/2$, we split the integral in three pieces. On the one
side, since $d\leq 3$ and since the maps $g$ and $y\mapsto |y|^{1/2}g(y)$
are in $L^1(\R_-)$,  we have
\begin{eqnarray*}
\left|\int_{-\infty}^{0}g(y)(1-\eps^{2/3}y)^{d/2-1}dy\right| &\leq&
\int_{-\infty}^{0}|g(y)|\max(2,2\eps^{2/3}|y|)^{1/2}dy=O(1).
\end{eqnarray*}
Then, there exists $C>0$ such that $g(y)\leq C(1+|y|)^{-1/2}$ for every $y\geq 0$, and since $d\geq 1$,
\begin{eqnarray*}
\left|\int_{0}^{\eps^{-2/3}/2}g(y)(1-\eps^{2/3}y)^{d/2-1}dy\right|\leq \int_{0}^{\eps^{-2/3}/2}C(1+|y|)^{-1/2}2^{1/2}dy=O(\eps^{-1/3}).
\end{eqnarray*}
Finally, 
\begin{eqnarray*}
\left|\int_{\eps^{-2/3}/2}^{\eps^{-2/3}}g(y)(1-\eps^{2/3}y)^{d/2-1}dy\right|
&=&\left|\int_{1/2}^{1}g(t/\eps^{2/3})(1-t)^{d/2-1}\frac{dt}{\eps^{2/3}}\right|\\
&\leq &C\eps^{1/3}\int_{1/2}^{1}t^{-1/2}(1-t)^{d/2-1}\frac{dt}{\eps^{2/3}}=O(\eps^{-1/3}).
\end{eqnarray*}
If $\alpha\geq 5/2$, using the Taylor formula,
\begin{eqnarray*}
\lefteqn{\int_{-\infty}^{\eps^{-2/3}}g(y)(1-\eps^{2/3}y)^{d/2-1}dy}\\ &=&
\int_{-\infty}^{+\infty}g(y)dy+\int_{-\infty}^{\eps^{-2/3}/2}g(y)\left((1-\eps^{2/3}y)^{d/2-1}-1\right)dy\\
&&+\int_{\eps^{-2/3}/2}^{\eps^{-2/3}}g(y)(1-\eps^{2/3}y)^{d/2-1}dy-\int_{\eps^{-2/3}/2}^{+\infty}g(y)dy\\
&=&
\int_{-\infty}^{+\infty}g(y)dy+\int_{-\infty}^{\eps^{-2/3}/2}g(y)\left(-(\frac{d}{2}-1)\eps^{2/3}y+(\frac{d}{2}-1)(\frac{d}{2}-2)\frac{1}{2}(1-\xi_{\eps,y})^{d/2-3}\eps^{4/3}y^2\right)dy\\
&&+\int_{1/2}^1g(t/\eps^{2/3})(1-t)^{d/2-1}\frac{dt}{\eps^{2/3}}+O((\eps^{-2/3})^{-5/2+1}),
\end{eqnarray*}
where $\xi_{\eps,y}\in [0,\eps^{2/3}y)$ or $\xi_{\eps,y}\in
  (\eps^{2/3}y,0)$, depending on the sign of $y$. In particular, if
  $y\leq \eps^{-2/3}/2$, $\xi_{\eps,y}\leq 1/2$ and since $d/2-3<0$,
  $(1-\xi_{\eps,y})^{d/2-3}\leq 2^{3-d/2}$. Additionnally, 
$$  \int_{-\infty}^{\eps^{-2/3}/2}g(y)y^2dy\leq\int_{-\infty}^{0}g(y)y^2dy+\int_{0}^{\eps^{-2/3}/2}C(1+y)^{-5/2}y^2dy=O(\eps^{-1/3})$$
   Moreover, there exists a
  positive constant $C$ such that for $t\in
  [1/2,1]$, $g(t/\eps^{2/3})\leq (\eps^{2/3}/t)^{5/2}$. Thus,
\begin{eqnarray*}
\int_{-\infty}^{\eps^{-2/3}}g(y)(1-\eps^{2/3}y)^{d/2-1}dy &=&
\int_{-\infty}^{+\infty}g(y)dy-(\frac{d}{2}-1)\eps^{2/3}\int_{-\infty}^{\eps^{-2/3}/2}yg(y)dy+O(\eps)\\
&=&
\int_{-\infty}^{+\infty}g(y)dy-(\frac{d}{2}-1)\eps^{2/3}\int_{-\infty}^{+\infty}yg(y)dy+O(\eps).
\end{eqnarray*}
If $\alpha\geq 3/2$, we start the calculation like in the case
$\alpha\geq 5/2$. Then, again thanks to the Taylor formula,
\begin{eqnarray*}
\left|\int_{-\infty}^{\eps^{-2/3}/2}g(y)\left((1-\eps^{2/3}y)^{d/2-1}-1\right)dy\right|
&=&\left|-(\frac{d}{2}-1)\int_{-\infty}^{\eps^{-2/3}/2}g(y)(1-\xi_{\eps,y})^{d/2-2}\eps^{2/3}ydy\right|\\
&\leq& \left|\frac{d}{2}-1\right|2^{2-d/2}\eps^{2/3}\int_{-\infty}^{\eps^{-2/3}/2}|g(y)||y|dy=O(\eps^{1/3}),
\end{eqnarray*}
for some $\xi_{\eps,y}\in [0,\eps^{2/3}y)$ or $\xi_{\eps,y}\in
  (\eps^{2/3}y,0)$. Then,
\begin{eqnarray*}
\int_{\eps^{-2/3}/2}^{\eps^{-2/3}}g(y)(1-\eps^{2/3}y)^{d/2-1}dy
&=&\int_{1/2}^1g(t/\eps^{2/3})(1-t)^{d/2-1}\frac{dt}{\eps^{2/3}}=O(\eps^{1/3}).
\end{eqnarray*}
Finally,
\begin{eqnarray*}
\int_{\eps^{-2/3}/2}^{+\infty}g(y)dy=O(\eps^{1/3}),
\end{eqnarray*}
which gives the result in the case $\alpha\geq 3/2$.

\paragraph{Aknowledgements.} I am grateful to Franco Dalfovo for
calling my attention to this problem, and to Amandine Aftalion for
helpful comments on preliminary versions of the manuscript.


\begin{thebibliography}{99}
\bibitem[AAB]{AAB} A. \textsc{Aftalion}, S. \textsc{Alama}, and
  L. \textsc{Bronsard}, \textit{Giant Vortex and the Breakdown of
    Strong Pinning in a Rotating Bose-Einstein Condensate},
  Arch. Rat. Mech. Anal., {\bf 178}, 247--286 (2005)

\bibitem[DPS]{DPS} F. \textsc{Dalfovo}, L. \textsc{Pitaevskii},
  S. \textsc{Stringari}, \textit{Order Parameter at the Boundary of a Trapped Bose Gas}, Phys. Rev. A 54, 4213-4217 (1996) 

\bibitem[F]{Fermi} E. \textsc{Fermi}, \textit{Statistical method of investigating
electrons in atoms}, Z. Phys. {\bf 48}, 73--79 (1928)

\bibitem[FIKN]{FIKN} A.S. \textsc{Fokas}, A.R. \textsc{Its},
  A.A. \textsc{Kapaev} and V.Y. \textsc{Novokshenov},
  \textit{Painlev\'e Transcendents, The Riemann-Hilbert Approach}, Mathematical Surveys and Monographs, 128. AMS, Providence, RI, 2006.

\bibitem[GP]{GP} C. \textsc{Gallo} and D. \textsc{Pelinovsky},
  \textit{On the Thomas-Fermi ground state in a harmonic potential},
  preprint. 

\bibitem[HM]{HM} S.P. \textsc{Hastings} and J.B. \textsc{McLeod}, \textit{A
  boundary Value Problem Associated with the Second Painlev\'e
  Transcendent and the Korteweg-de Vries Equation},
Arch. Rat. Mec. Anal., {\bf 73}, 31--51 (1980)

\bibitem[IM]{IM} R. \textsc{Ignat} and V. \textsc{Millot}, \textit{The critical velocity
for vortex existence in a two-dimensional rotating Bose--Einstein
condensate}, J. Funct. Anal. {\bf 233}, 260--306 (2006)

\bibitem[M]{M} D. \textsc{Margetis}, \textit{Asymptotic formula for
  the condensate wave function of a trapped Bose gas}, Phys. Rev. A
  {\bf 61}, 055601 (2000)

\bibitem[T]{Thomas} L.H. \textsc{Thomas}, \textit{The calculation of atomic
fields}, Proc. Cambridge Philos. Soc. {\bf 23}, 542 (1927)


\end{thebibliography}
\end{document}